# Power Loss Minimization of Distribution Network using Different Grid Strategies

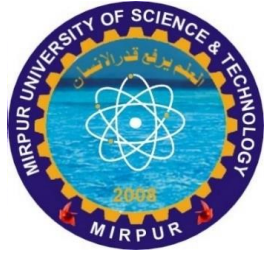

**UMAR JAMIL**
**(1JZ22100O3100517)**

Session 2015-2018

**Department of Electrical Engineering**
**Faculty of Engineering**
**Mirpur University of Science and Technology**
**Mirpur, (AJK),**
**Pakistan.**

# Power Loss Minimization of Distribution Network using Different Grid Strategies.

**By**

**UMAR JAMIL**
**(1JZ22100O3100517)**

A Thesis submitted in partial fulfillment of
the requirements for the degree of

# MASTER OF SCIENCE

# IN

# ELECTRICAL ENGINEERING

Session 2015-2018

**Department of Electrical Engineering**
**Faculty of Engineering**
**Mirpur University of Science and Technology**
**Mirpur, (AJK),**
**Pakistan.**

# CERTIFICATION

I hereby undertake that this research is an original one and no part of this thesis falls under plagiarism. If found otherwise, at any stage, I will be responsible for the consequences.

Student's Name: <u>Umar Jamil</u>          Signature: 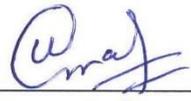

Registration No.: <u>1JZ22100O3100517</u>          Date: <u>O9/O4/18</u>

Certified that the contents and form of thesis entitled "Power Loss Minimization of Distribution Network using Different Grid Strategies." submitted by **"Mr. Umar Jamil"** have been found satisfactory for the requirement of the degree.

Supervisor: 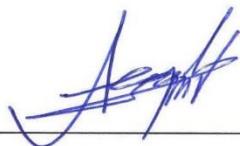

External Examiner: 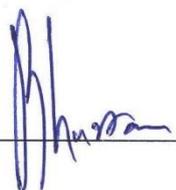

Chairperson: 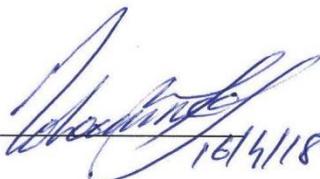

Dean: 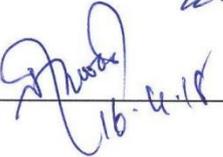

Director Advance Studies
& Research Board: 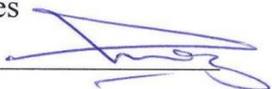

Director
Advance Studies & Research
Mirpur University of Science &
Technology. (MUST) Mirpur

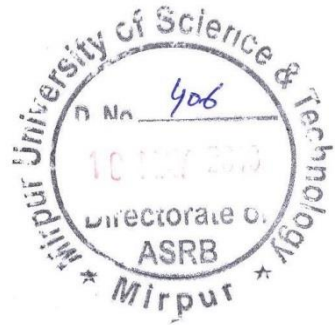

iii

The presented work in this thesis is dedicated to my

**Grand Parents,**

**Parents & Teachers.**

**Table of Contents**

# List of Figures

# List of Tables

# Abbreviations

AC ...................................................................................................... Alternating Current

DGs ............................................................................................... Distributed Generations

FDIR ...................................................................... Fault Detection Isolation & Restoration

GS ............................................................................................................ Golden Search

GA ............................................................................................................ Genetic Algorithm

GPSO-GM ...  Guaranteed convergence Particle Swarm Optimization with Gaussian Mutation

IGSA-CSS  ......... Improved Gravitational Search Algorithm-Conditional Selection Strategies

IBDGs ...................................................................................................... Inverter Based DGs

KCL .................................................................................................. Kirchhoff's Current Law

KVL .................................................................................................. Kirchhoff's Voltage Law

NR ............................................................................................................ Newton Raphson

OPF ...................................................................................................... Optimum Power Flow

PSO ...................................................................................................... Particle Swarm Optimization

PV  ............................................................................................................ Photovoltaic

PG ...................................................................................................... Real Power Generation

# ACKNOWLEDGMENT

Thanks to ALLAH ALMIGHTY WHO always help me in every part of life and to complete this research work and MSc degree. Thanks to my beloved parents who supported me throughout my study and life. My sincere appreciation and gratitude are paid to my supervisor Sir Dr. Anzar Mahmood (Associate Professor, EPED MUST AJK), for his guidance, support and encouragement throughout my thesis work. I am also very thankful to Engr. Adil Amin and Engr. Salman Bari for their cooperative response in this research work.

**Umar Jamil**

# ABSTRACT

Power losses in electrical power systems especially, distribution systems, occur due to the several environmental and technical factors. Transmission & Distribution line losses are normally 17% and 50% respectively. These losses are due to inappropriate size of conductor, long distribution lines, low power factor, overloading of lines etc. The power losses cause economic loss and reduce the system reliability. The reliability of electrical power system can be improved by decreasing network power loss and by improving voltage profile. In radial distribution system, power loss can also be minimized through Distributed Generation (DGs) system placement. In this thesis, three different grid strategies including real power sharing, reactive power injection and transformer tap changing are discussed and used to minimize line losses. These three proposed grid strategies have been implemented using power flow study based on Newton Raphson (NR) and Genetic Algorithm (GA). To minimize line losses, both methods have been used for each grid strategy. The used test system in this research work is IEEE-30 bus radial distribution system. Results obtained after simulation of each grid strategy using NR and GA shows that real load sharing is reliable with respect to minimization of line loss as compare to reactive power injection and transformer tap changing grid strategy. Comparative analysis has been performed between GA and NR for each grid strategy, results shows that Genetic Algorithm is more reliable and efficient for loss minimization as compare to Newton Raphson. In base case for optimum power flow solution using genetic algorithm and newton Raphson, real line losses are 9.481475MW and 17.557MW respectively. So, GA is preferable for each proposed grid strategies to minimize line losses than NR.

# INTRODUCTION

## 1.1 CONVENTIONAL POWER SYSTEM

Power Generation, Transmission and Distribution are main parts of electrical power system. Delivering of efficient and lossless power to consumers is a big challenge for the electric utilities. Reliability is a main factor which is considered for efficient operation of power system. Reliability of power system means continuity of power supply without any interruption. Generated power must be delivered to consumer through transmission and distribution systems [1]. Conventional electrical power system is shown in figure 1.1.

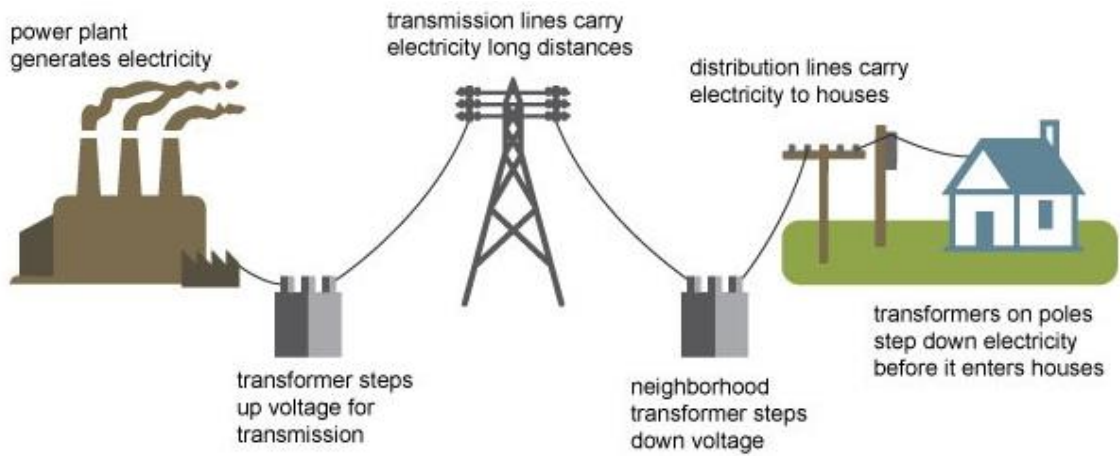

Fig. 0.1: Conventional Electrical Power System [2]

Power plant generates electricity and then power transformers step up the generated voltage. After step up, high voltages are transmitted through transmission lines that are used for long distances and near the consumer, high voltages are step down through substation transformer for utility purpose. Only one third of fuel energy is converted into electricity in conventional existing grid because of its unidirectional nature. In addition, 8% of output power is wasted along lines and 20% of generated capacity is required for peak demand [3].

## 1.2 DISTRIBUTION SYSTEM

The part of electrical power system which is used to distribute power to consumers through distribution lines is termed as distribution system. Distribution system mainly composed of feeder distributer and service mains and lies in generally between transmission and consumer meters. The area where power needs to be distributed is connected with substation through a conductor known as feeder. Tapings for consumer supply are taken from conductor known as distributor. Connection between distributor and consumer terminal is done by small cable known as service mains. Main consideration of designing of feeder and distributer is current carrying and voltage drops along lines respectively. In distribution network, voltage variation is allowed up to ± **6%** of consumer terminals rated value. Main requirements of distribution systems are reliability, proper standard voltage level and power availability on demand. Distribution system can be classified in to radial, ring main and interconnected systems according to connection's scheme [4]. Radial distribution system is used in this thesis. The classification of power distribution system is shown in Figure 1.2.

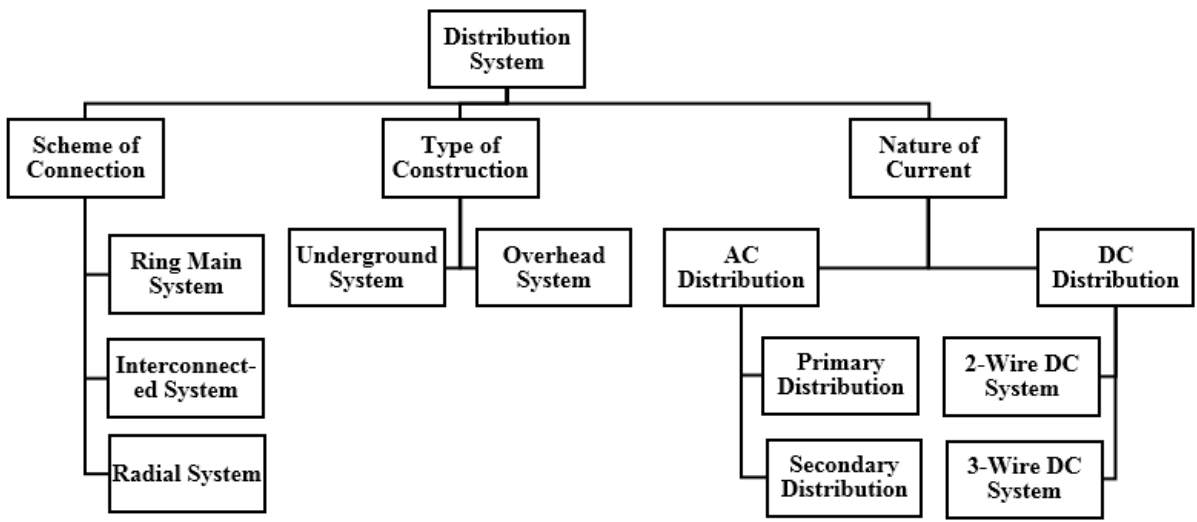

Fig. 0.2: Classification of Power Distribution System

## 1.2.1 Radial Distribution System

The distribution system has a separate feeder emitted from substation is used for distributors feeding at one end only is termed as Radial distribution system. Conventional radial distribution system is going to change to ring main and interconnected system due to electrical system deregulation and DGs connection at medium and low voltage. But conventional radial distribution has advantages of voltage regulation, reduction of loss, increase of reliability and facilitating use of DGs [5]. Single line diagram of AC radial distribution system is shown in Figure 1.3 [6]**.** Three phase unbalanced radial distribution feeders can be analyzed through basic data like load models, Shunt Capacitors, Overhead spacing models, conductor data, underground spacing models, cable data, configuration codes, line segment data and voltage regulators [7].

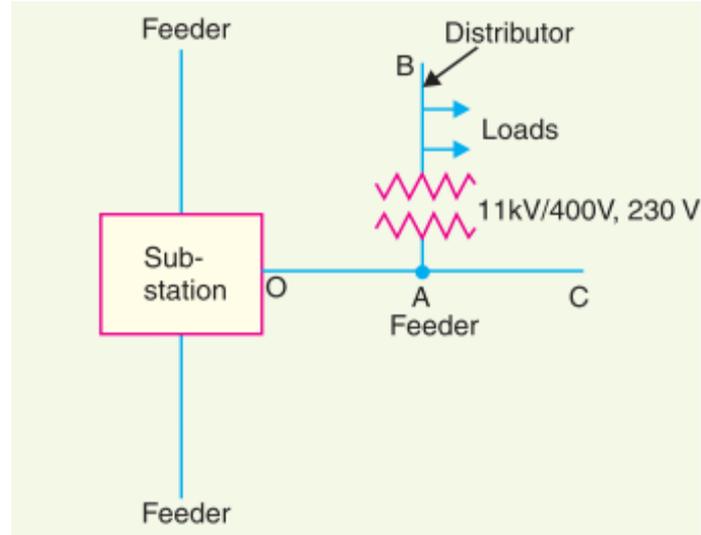

Fig. 0.3: One Line Diagram of Radial Distribution System

Commonly used radial distribution test system for power flow solution are IEEE 14, IEEE-30, IEEE-33 IEEE 67and IEEE 123. In this thesis, IEEE-30 BUS system is used as test case to determine and minimize power losses, as a result voltage profile is improved.

## 1.3  POWER FLOW ANALYSIS

Power flow or load flow studies are used to determine active power, reactive power, voltage, and power factor, current and power loss of the system. It can also determine location and size of DGs and capacitors. The backbone of power system is load flow analysis. Phase angle and voltage magnitude can be obtained through load flow analysis [8]. In addition, grid strategies including real load sharing, reactive power injection and transformer tap changing can be used to reduce line losses. In case of real load sharing strategy, sharing of load from highly loaded bus to low loaded bus can minimize line losses. Injection of reactive power at low voltage bus can improve voltage profile and reduce line losses. In case of tap changing grid strategy, through tap changing line losses can be minimized.

## 1.4  CHALLENGES OF DISTRIBUTION SYSTEM

Radial distribution system is normally simple in design having various challenges which are following presented in [9], [10].

1. Different phases like three phases, single phase or two phase of line sections of distribution system

2. Load Distribution: Load distribution is unbalanced usually on distribution feeder

3. Limited data is required for modelling and analysis due to less monitoring and control

4. High power loss and reduction of voltage due to insufficient support of reactive power in distribution network

5. Comprehensive study is required for DG and capacitor placement which have capability to supply reactive power

## 1.5 POWER DISTRIBUTION & TRANSMISSION LINE LOSSES

Difference of generated and distributed units is known as power distribution and transmission line loss. These are not paid by the consumers. Transmission line losses are approximately 17% while distribution line losses are approximately 50% [11]. In addition, there are two major types of line losses for power transmission and power distribution systems which are technical and commercial losses. Normally technical losses are 22.5%. Main reason of technical losses of distribution line is following:

a) Inadequate size of distribution line conductors

b) Lengthy power distribution lines

c) Distribution transformer are not installed at load centers

d) Due to less power factor of primary distribution system and secondary distribution system

e) Overloading of lines

Commercial or non-technical losses are 16.6% and main reason of these losses are following:

a) Power theft

b) Inaccuracies of meters

c) Error in meter reading

d) Losses of un-metered for low loads

e) Un-metered supply

## 1.6 POWER LOSS MINIMIZATION TECHNIQUES FOR DISTRIBUTION NETWORK

Power losses can be reduced through placement of DGs, injection of reactive power, transformer tap changing and real load sharing [12]. DG technologies are classified into 4 main divisions: Injection of active and reactive power through DG, Injection of only active or real power through DG, Injection of reactive power through DG, capability of DG to inject real power and consume reactive power. Through optimal placement of sizing of DG and capacitor, power losses can be reduced [10]. In this thesis, three grid strategies are used to minimize power losses and to improve voltage profile. These grid strategies are real load sharing from highly loaded bus to low loaded bus, Injection of reactive power at low voltage bus and transformer tap changing. Theses grid strategies are implemented on IEEE-30 Bus test system using Newton Raphson load flow analysis and optimum power flow analysis using Genetic Algorithm. Table 1.1 shows losses at different distribution stages [13].

Table 0.1: Losses at Distribution Stages

| S.No. | Component of Distribution System | Estimated Loss as a %age of Energy Sold | |
|---|---|---|---|
| | | **Rural Area** | **Urban Area** |
| 1 | Sub transmission Lines | 0.1 | 0.7 |
| 2 | Power Transformers | 0.1 | 0.7 |
| 3 | Distribution Lines | 0.9 | 2.5 |
| 4 | Distribution Transformer at No Load | 1.2 | 1.7 |
| 5 | Distribution Transformer at Load | 0.8 | 0.8 |
| 6 | Secondary Lines | 0.5 | 0.9 |
| 7 | **Total** | **3.6** | **7.3** |

## 1.7 OBJECTIVES

The objectives of this research are

i) To study and understand different techniques of power loss minimization
ii) To implement the grid strategies to minimize real line losses of distribution network using N-R and GA
iii) Comparison of implemented grid strategies using NR & GA with respect to real loss minimization

## 1.8 THESIS ORGANIZATION

Power loss minimization techniques for power distribution system are presented in this research work. An improved voltage profile is obtained in results. The work in the thesis is arranged as: Chapter 2 is related to literature review of previous proposed strategies for loss minimization and to increase reliability of distribution system. Chapter 3 discuss the proposed methodology while simulations and results are discussed in chapter 4. In chapter 5 conclusion and future recommendations are described.

# REVIEW OF LITERATURE

Power Flow analysis plays important role in electrical distribution network system. This chapter deals with the literature review of various schemes that have been implemented previously to minimize line losses.

In [14], authors presented a fast decoupled power flow method for unbalanced radial power distribution systems. Less number of equations are required and structural and numerical properties of distribution systems can be exploited through proposed method. Three phase PFS technique on the basis of propagation of forward and backward was proposed by authors in [15]. The technique is preferable to operation of real time of distribution network. This technique is used to analyze large practical distribution feeders. The efficient proposed method is applicable to compensation of reactive power and reconfiguration of network.

In [16], authors presented an most effective backward /forward load flow analysis method for radial distribution systems. The method behavior was investigated for different models of loads in detail. The convergence is not affected through the power factor of load and by the ratio of X/R and it mainly depends on the load admittance and line impedance equivalent magnitude. Authors suggested power grid operations and actions with variance time scale up to years from microseconds [17]. Telecommunications, power grid, cost of cascading failures, proposed agent technology and MAG systems are discussed in this paper. Vulnerability and network reliability increases due to self-healing infrastructure in electrical power grid.

For selection of optimal conductor to support voltage of feeder, a reliable computer algorithm was presented in [18]. Minimization of energy loss cost and conducting material's capital cost of depreciation and interest were obtained through proposed algorithm. Load flow method on the basis of interval arithmetic for radial distribution system was presented in [19]. It was concluded by authors that in the initial stage of any design and planning study at least, the presented method can be useful. The obtained solution through arithmetic interval-based method includes all the obtained solutions from simulations repeatedly.

 A useful simulation annealing technique of optimization to solve the OPF is presented by authors in [20]. The proposed technique is suitable because there is no need to calculate equations of differentials and Jacobean matrix as compare to classical methods. OPFSA constraints parameters includes losses of systems, technical limits of generators, limits of line and static indices security. Large computational time is the drawback of proposed method to achieve optimal solution. Problem of reactive scheduling can be solved through this method.

On the basis of backward/forward sweep load flow analysis, connections of various windings of distribution transformers can be handled for radial unbalanced power distribution systems. The authors concluded that existing matrices of nodal admittances of power distribution transformers can be implemented efficiently through proposed method. Accurate shifting of phase produced due to different connections of transformer winding and phase to ground equivalent voltages which are automatically produced can be solved by resulting algorithms [21]. In [22], authors presented an optimal reconfiguration method of radial distribution systems. The proposed method was formulated on the basis of maximum index of loading. To obtain best profile of voltage and maximization of margin of loadability, method of fuzzy modelling was used. Authors implemented this proposed method on standard IEEE-33 bus radial distribution system.

For optimal location determination to place DG at radial systems, authors presented an analytical method in [23]. Through this method power loss can be minimized. It is not iterative like power flow or load flow algorithms. There are no convergence problems and very quickly results can be obtained. In [24], authors presented an efficient, simple and fast convergence power flow method which was based on the forward and backward voltage .This techniques was applied to 12 and 30 bus system without laterals and with laterals respectively. Less number of iterations are required and it is less sensitive to the parameters of distribution system. Authors proposed an efficient heuristic algorithm for switches sensitivity determination at open status using OPF solution in [25]. Obtained results determine that there were reduction of power flows and performance of proposed method had been enhanced and reliable for real distribution networks.

For three phase power flow analysis, an improved algorithm of backward/forward sweep was presented in [26]. For each transformer branch and line, KCL and KVL are used in the backward sweep. The procedure of solution terminates if voltage tolerance is greater than mismatch at the substation. The proposed method is accurate and better than other two mostly used methods.

Authors presented a deep small voltage stability disturbance of power system through combination of dynamic, static, nonlinear and linear tools of analysis in [27]. They introduced ill and well condition of stability behavior. Small voltage disturbance stability and sensitivities according to the states of power system under different conditions are discussed and presented.

To improve voltage profile and maximization of net savings of distributions system, optimal placement of capacitor using multi-objective GA based fuzzy logic approach has been presented in [28].

Equations of systems which describes radial power distribution system were presented by authors in [29]. By using I[st] order Newton Raphson method, set of equations can be solved. The proposed Newton Raphson method remain in stable condition even when radial distribution system is approaches to near its maximum point of loading. This method is used to help in computation of the maximum point of loading or voltage point of collapse. The minimum value of singular's behavior was also discussed in the absence and presence of capacitors. Capacitors' effects on the system singularity had also been discussed. To assess the particular load condition stability, authors proposed a new voltage stability method for radial distribution system. Load-ability can be increased due to power factor improvement at specific bus. Two bus equivalent system slightly changed in this case [30].

In [31], authors presented major contribution on modification of IEEE 6,14 and 30 systems. They computed optimized weighting factor's value and obtained optimal size and locations of Distributed Generations (DGs). It was observed due to that load voltages are improved and losses are minimized. The obtained results encourage the proposed strategy implementation in a distribution network on large scale. To connect consumers to a radial distribution system, loss allocation method was proposed in [32]. Comparison of proposed exact method with Quadratic loss allocation and pro rata is presented in detail. Pro rata method is easy and simple for implementation and same power loss obtained having equal load demands to allocate consumers. Method of Quadratic loss allocation indicates that the factor of loss allocation is proportional to square of active or reactive load current for a particular consumer.

Authors presented a linear programming method for solution of allocation problem of voltage regulators capacitors in radial power distribution systems in [33]. Calculations of power losses and magnitude of voltages with high precision were obtained as compare to load flow sweep method.

In order to perform optimum control action of self-healing smart grid, comprehensive design with operating strategy were introduced. System is divided into two main stages. In planning stage, Active electrical distribution system was divided into multiple micro grids with DG units and load's probabilistic nature consideration. In operating stage, self-healing control actions presented probabilistically. Reconfiguration of system, load shedding and output of DGs are included in control actions. Optimal control strategy depends on any one of these 3 factors, self-adequacy in the micro grids having no fault, total energy losses and sum of all loads supplied by system. In order to achieve more reliable and intelligent smart grid, self-healing control actions will be suggested [34].

To solve the problems of DGs like size, allocation, type in radial distribution systems, authors presented an approach of mixed integer linear programming. Minimization of annual investment and cost of operations was the objective function. Through obtained results of test system, accuracy and efficiency of respective proposed method was achieved [35]. In [36], authors proposed a fast combined technique for solution of location and size of DGs. Size calculation and location determination were done through loss sensitivity factor and simulated annealing respectively. For loss minimization and voltage profile improvement, developed method was implemented on large scale 33 bus, 9 bus and 118 bus radial distribution system. It was concluded by authors that developed technique can be implemented for any system size with capability to solve DGs size problem and optimal DGs placement.

To handle PV and PQ buses, authors introduced a new algorithm of radial load flow in [37]. With the use of quadratic equation solution, the method was developed on the basis of direct power flow solution achieve from a system of two buses. The proposed method was tested on two renewable energy-based DG radial systems. The Quadratic convergence of proposed method in the PV bus presence was indicated through results. In [38], authors presented method of placement of multi DG units and its sizing based on load ability of systems; maximization. They proposed hybrid particle swarm optimization and then compared results with Etihad method. They concluded after comparison that proposed method is reliable and efficient with respect to power losses reduction and voltage profile improvement.

In radial distribution network systems, energy storage size and placement methodology optimally and practically were addressed in [39]. The aim of proposed methodology to increase the system reliability. Mixed integer nonlinear programming formulation was used for problem and solved through PSO algorithm. The obtained results show the reduction of network operation cost and improved reliability. In [40], authors developed a new voltage stability index and two types of distributed generations are used. At unity and at 0.9 lagging power factor, operation of DGs are performed and it was concluded that real and reactive power loss reduction and voltage profile improvement are more significant in case of DG having lagging power factor as compare to DG at power factor of unity.

In [41], authors discussed that the pattern of operation in electrical distribution system, radial topology is the less effective if there is system goal for minimum power system losses operation. Quasi radial system topologies having power losses are very close to minimum loss of power. With quasi radial topology, Operation of distribution systems reduces the loss of power from the system. Bat algorithm was used on IEEE-34 bus distribution system to minimize power losses [42]. For pre-identification of capacitor optimal location, factor of loss sensitivity has been used. From the proposed method, net savings, minimum compensation locations as compare to other available techniques, and maximum reduction of power loss were achieved. Stability of power system improves due to decrease in MVA intake from grid. Voltage profile enhancement and of power loss minimization are achieved from proposed method.

Improvement of voltage profile and real and reactive losses reduction due to placement of DG were observed using Newton Raphson method based on power flow. In this paper authors implemented the proposed technique on different 3 IEEE bus systems. To deliver real power only 2MW capacity of photovoltaic cell as DG was used in this study [43]. In [44] authors proposed a whale optimization algorithm for capacitor sizing and location in distribution network system. Due to capacitor placement at optimal points of location, stability with system performance and reliability will be improved. Radial distribution IEEE-34 and IEEE-8 bus test system was used on which proposed method was implemented.

To find location and optimum number of remote control switch in a radial feeder of distribution system, authors implemented differential search algorithm in [45]. It was observed after comparison of proposed algorithm with other algorithm that proposed algorithm's performance is better and efficient method for solution of multi objective remote-control switch placement issues. In [46] authors proposed an IGSA-CSS and GSA-CSS algorithms for power loss minimization and voltage profile improvement through optimal reactive power control. Proposed method was implemented on IEEE 30 and IEEE 57 test bus systems. It was observed by authors that IGSA based algorithm is more accurate and effective optimal solution can be obtained in high quality.

In [47], authors presented a novel approach to control voltage at the level of MV. The research is based on analysis of day-ahead that utilize data from forecasting of load or renewable energy resources for establishment of next day plan for different distributed energy resources operations by Multi-temporal OPF approach. Authors in this paper have discussed detailed review of power flow analysis of different DGs modes. For minimization of total active power mismatch, PSO were used. GPSO-GM, a novel algorithm is also proposed in this paper for power flow analysis problem in a micro island grid. The proposed algorithms were implemented on 6, 33 and 69 bus systems. Best results have been d from proposed algorithm [48].

In [49], authors proposed an improved model of network with network and reactive power loss consideration. Voltage magnitude influence is included in loss term. Combined mathematical transformation and Taylor expansion series is used. As compare to other existing network models, proposed model's accuracy is improved. The formulation of OPF model is presented using proposed network model. A convex relaxation algorithm is introduced due to introduction of non-convexity of the model of optimum power flow analysis through the term of quadratic loss. Several benchmark systems of IEEE are used for demonstration of desirable proposed model's performance.

Authors presented power losses for the whole day and for the system's steady state condition with the consideration of load profiles. Neplan software was used for determination of bus voltages of buses ,generation of real power through the generators, consumption of real power through the loads and the power losses using Newton Raphson method [50]. This paper presented efficient research on optimum sizing and placement of DGs in radial distribution systems. Four different cases were discussed including base case, real power DG, reactive power DGs and combined real-reactive power DG. Results obtained from combined real reactive power DG were more efficient for optimum radial distribution system, Voltage profile has been improved and losses are reduced as compared to other cases [51].

Protective devices operation may be affected through contribution of inverter based DGs (IBDGs) fault current. So short circuit analysis approach were presented by authors in [52]. They concluded that for load, the methodology developed in this study is more accurate and have capability to include IBDGs voltage control dependent modes. The source current deviation and increment in fault current will increase the IBDGs penetration.

Stability of voltage index for electrical power distribution systems was presented in [53]. Distributed way and scalable computation can be performed for voltage stability index approximately. The proposed approach can be contributed significantly in online tool of voltage stability monitoring, in optimal programming of power flow and in numerically algorithm in power flow construction. The proposed method can be used to improve large networks scalability and to avoid exorbitant computations determinant.

To solve the problem of the non-convex AC OPF in radial distribution networks for convergence of local minima, authors proposed an algorithm based on Lagrangian augmented approach that depends on multipliers method. The proposed algorithms are used to solve formulations of decentralized and centralized OPF targets. Robustness of version of centralized was shown with respect to various lengths of line, rated network voltage values, different operating points of networks, shunt capacitors presence in the grid and different states of electrical systems' initial conditions [54].

Authors proposed an OPF for supervision of various power distribution networks [55]. Voltages of high voltage network can be supported by the medium voltage grid through exchange control of reactive power at the both high and medium voltage grid interface. Investigations were performed between the power distribution grids interaction with optimal coordination and VOLT VAR controllers.

In [56], authors proposed a recursive power flow algorithm to balance or unbalance radial power distribution systems. For power flow solution, on the basis of graph theory four constant matrixes are used. Convergence and solvability characteristics mathematically presented through proposed method.

Uncertainty of DG output and load impact on bus voltages of power distribution network can be studied by proposed power flow method such as multi stage affine arithmetic in [57]. IEEE 123 bus test system was used and authors concluded that proposed method can obtain accurate upper and lower bound bus voltage with fast computational speed.

In the next chapter of this thesis, methodology to minimize real line losses using N-R and GA is discussed. Control grid strategies which are implemented on IEEE-30 Bus radial distribution system are discussed in next chapter.

# METHODOLOGY

## 3.1 POWER FLOW STUDY

An essential tool for power system involving numerical simulations in steady state normal operation is termed as power flow or load flow study. Per unit system and single line diagram are used in power flow study. Both real and reactive power are under consideration. In this chapter, proposed methodology is discussed.

Power flow study has following merits in power system. Line losses, system voltage profiles, stability and active and reactive power flow information are provided by power flow study in operation. Future analysis in project development are provided by this load flow study like new power generation station and additional unit, new power transmission and power distribution lines, load forecasted demand and also interconnection of new systems with already existing systems. Newton Raphson and Genetic Algorithm are used in this thesis for loss minimization through real load sharing, reactive power injection and transformer tap changing grid strategies.

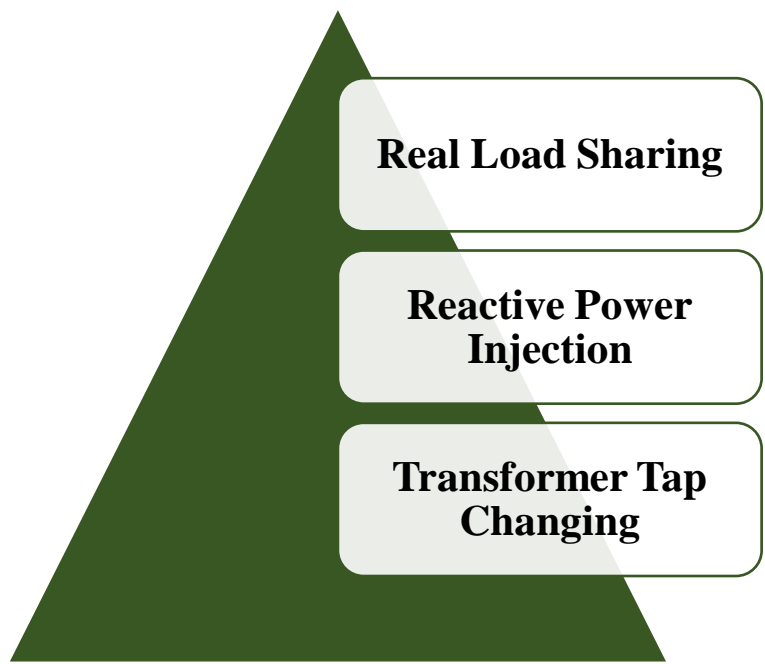

Fig. 0.1: Grid Strategies to Minimize Line Losses

## 3.2 FLOW CHART OF METHODOLOGY

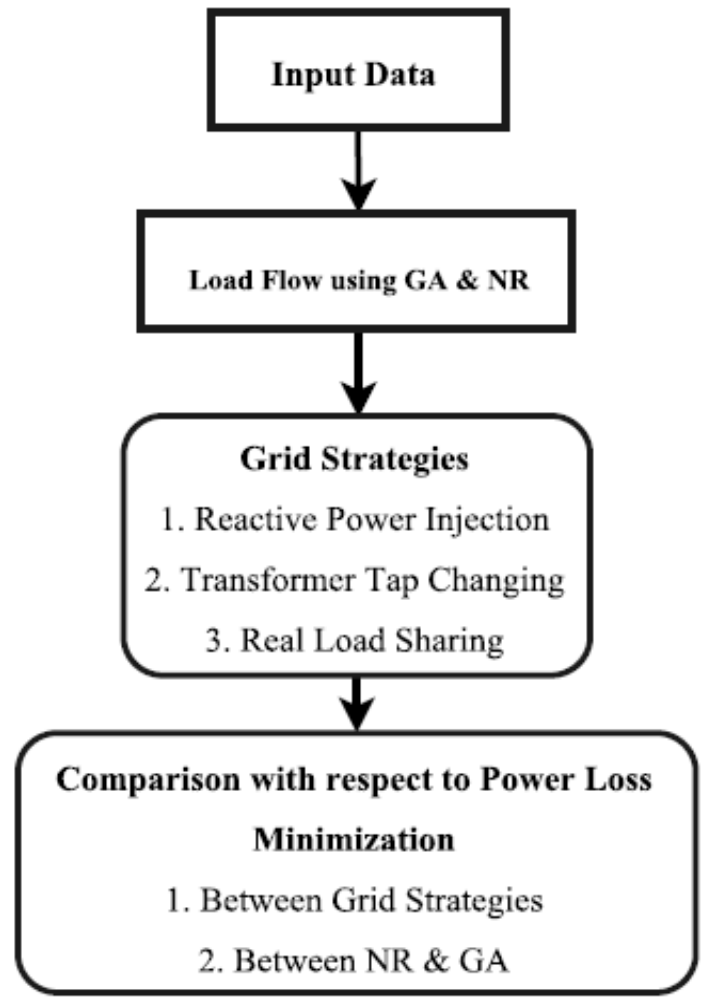

Fig. 0.2: Flow Chart of Methodology for Loss Minimization

## 3.3 EQUATIONS FOR LOAD FLOW SOLUTION

There is need to define the following parameters from load flow equations to formulate the real and reactive power which enters to a bus [12].

### 3.3.1 Bus Admittance Matrix ($Y_{bus}$)

$Y_{bus}$ is used to relate injected current at a bus with the bus voltages. According to Kirchhoff's current law (KCL), sum of current injected to a node is equal to sum of leaving current from node i.e total current in a closed loop will be zero. It is fundamental tool for network analysis and can be formulated by node voltage equation I= $Y_{bus}$V [58]. Where I and V are vector injected node current and vector of node voltage respectively. Bus Admittance matrix has following categories:

- ➢ **Symmetric Matrix:** $Y_{bus}$ (m, k)= $Y_{bus}$(k, m)
- ➢ **Diagonal Entries:** $Y_{bus}$ (m, m) is actually sum of all node J components admittance
- ➢ **Off Diagonal Entries**: $Y_{bus}$ (m, k) represents negative admittance of all those components which are connected between node J and I. Where m represents bus admittance of node J component and k represent bus admittance of node I components.

**Voltage at I$^{th}$ Bus**

$$V_I = |V_I| \angle \alpha_I = |V_I|(\cos \alpha_I + j \sin \alpha_I) \tag{4.1}$$

Where $V_I$ is the voltage at I$^{th}$ bus and $\alpha_I$ is the voltage angle.

**Self-Admittance at I$^{th}$ Bus**

$$Y_{II} = |Y_{II}| \angle \theta_{II} = |Y_{II}|(\cos \theta_{II} + j \sin \theta_{II}) \tag{4.2}$$

Where $Y_{II}$ is the self-admittance at I$^{th}$ bus and $\theta_{II}$ is the self-admittance angle.

$$Y_{II} = G_{II} + jB_{II} \tag{4.3}$$

**Mutual-Admittance between Bus I and J**

$$Y_{IJ} = |Y_{IJ}| \angle \theta_{IJ} = |Y_{IJ}|(\cos \theta_{IJ} + j \sin \theta_{IJ}) \tag{4.4}$$

Where $Y_{IJ}$ is the mutual-admittance between I$^{th}$ and J$^{th}$ bus while $\theta_{IJ}$ is the mutual-admittance angle.

$$Y_{IJ} = G_{IJ} + jB_{IJ} \tag{4.5}$$

Suppose power system is consisting of n buses.

**Injected Current at I$^{th}$ Bus**

$$I_I = Y_{I1}V_1 + Y_{I2}V_2 + \cdots + Y_{IN}V_N \tag{4.6}$$

$$I_I = \sum_{K=1}^{N} Y_{IK}V_K \tag{4.7}$$

Where

$$V_K = |V_K| \angle \alpha_K$$

$$Y_{IK} = |Y_{IK}| \angle \theta_{IK}$$

Here $\alpha_k$ is the voltage angle for $V_K$ and $\theta_{IK}$ is the $Y_{IK}$ admittance angle

**Injection of Real and Reactive Power at Node I**

$$S_I = P_I + jQ_I = V_I I_I^* \tag{4.8}$$

$$S_I = \sum_{K=1}^{N} \lceil |Y_{IK}V_IV_K| \angle(\theta_{IK} + \alpha_I - \alpha_K) \rceil \tag{4.9}$$

Where

$$P_I = Re(S_I)$$

$$Q_I = Im(S_I)$$

**Complex Power at I$^{th}$ Bus**

$$P_I - jQ_I = V_I^* I_I = V_I^* \sum_{K=1}^{N} Y_{IK}V_K \tag{4.10}$$

$$= \sum_{K=1}^{N} \lceil |Y_{IK}V_IV_K|(\cos\alpha_I - j\sin\alpha_I)(\cos\theta_{IK} + j\sin\theta_{IK})(\cos\alpha_K + j\sin\alpha_K) \rceil \tag{4.11}$$

Load flow equations in terms of real and reactive power are following:

$$P_I = \sum_{K=1}^{N} \lceil |Y_{IK}V_IV_K| \cos(\theta_{IK} + \alpha_K - \alpha_I) \rceil \tag{4.12}$$

$$Q_I = -\sum_{K=1}^{N} \lceil |Y_{IK}V_IV_K| \sin(\theta_{IK} + \alpha_K - \alpha_I) \rceil \tag{4.13}$$

Total injected real power at I$^{th}$ bus

$$P_{I,inj} = P_{GI} - P_{LI} \tag{4.14}$$

The purpose of load flow analysis is to reduce or minimize the following real and reactive mismatches.

$$\Delta P_I = P_{I,inj} - P_{I,calc} = P_{GI} - P_{LI} - P_{I,calc} \tag{4.15}$$

$$\Delta Q_I = Q_{I,inj} - Q_{I,calc} = Q_{GI} - Q_{LI} - Q_{I,calc} \tag{4.16}$$

### 3.3.2 Line Losses

For nominal $\pi$ line representation, the power flow or line loss $P_{LI}$ at I$^{th}$ bus is calculated by

$$P_{LI} = P_{PQ} - jQ_{PQ} = V_P^* I_{PQ} \tag{4.17}$$

Where

$$I_{PQ} = \left[V_P - V_Q\right]Y_{PQ} + V_P \frac{Y'_{PQ}}{2} \tag{4.18}$$

## 3.4 OBJECTIVE FUNCTION

The function which must be minimized under different constraints is objective function. In this research work, minimization of real power loss is main objective or fitness function.

$$OF = FF = \min(P_{TLL}) \tag{4.19}$$

Where $P_{TLL}$ is the total real line losses of the system. FF is the fitness function which is objective function(OF) in genetic algorithm problem. Mathematically objective function for real line loss minimization can be calculated using following equations:

$$OF = \sum_{I,J}^{N} [P_I - P_J] \tag{4.20}$$

$$OF = \sum_{I=1}^{N} \sum_{J=1}^{N} \left( \frac{R_{IJ} \cos(\alpha_I - \alpha_J)}{V_I V_J} \left(P_I P_J + Q_I Q_J\right) + \frac{R_{IJ} \sin(\alpha_I - \alpha_J)}{V_I V_J} \left(Q_I P_J + P_I Q_J\right) \right) \tag{4.21}$$

Where, Line resistance between $I^{th}$ and $J^{th}$ bus is represented by $R_{IJ,}$ voltage and angle at $I^{th}$ bus is represented by $V_I$ & $\alpha_I$, number of Bus is represented by $N$ and total Injected real and reactive power at Ith bus is denoted as $P_I$ and $Q_{I.}$

## 3.5 BUS CLASSIFICATION

Following are 4 types of buses in electrical power system.

   i)     Slack Bus (Swing Bus)

   ii)    Generator Bus (PV Bus)

   iii)   Load Bus (PQ Bus)

   iv)    Voltage Controlled bus

Following are 4 variables in power system

   i)     Real Power (P)

   ii)    Reactive Power (Q)

   iii)   Voltage Magnitude (V)

   iv)    Voltage Angle ($\delta$)

Table 0.1: Classification of Power System Buses

| No. | 1 | 2 | 3 | 4 |
|---|---|---|---|---|
| Bus Types | Slack/Swing bus | Generation/PV bus | Laod /PQ bus | Voltage Controlled bus |
| Known Variables | $|V|, \alpha$ | $P_g, |V|$ | $P_g, Q_g$ | $P_g, Q_{g,} |V|$ |
| Unknown Variables | $P_g, Q_g$ | $Q_g, \alpha$ | $|V|, \alpha$ | $\alpha$, a |
| Remarks | $|V|, \alpha$ will be assumed if not specified as 1 and 0 deg | Generator will be present at machine bus | Approx. 80% buses are of PQ type | Where a is the % tap change in tap changing T/F. |

## 3.6   DATA FOR LOAD FLOW

Load flow or power flow data is required for load flow analysis. Formulation of bus admittance matrix is possible from this data. Necessary data for load flow is as under [59]:

### 3.6.1   System Data

Number of buses(n), PV buses, transmission lines load, shunt elements transformers, the number of slack bus, slack bus voltage magnitude, slack bus angle is taken as 0-degree, tolerance limit, base MVA, possible maximum number of iterations are included in the system data.

### 3.6.2   Generator Bus Data

Bus number, Real power generation PG, the specified magnitude of voltage, minimum and maximum limit of reactive power are included in generator bus data.

### 3.6.3   Load Data

The bus number, real and reactive power demand data are required for all loads.

### 3.6.4   Transmission Line Data

The bus starting number I, bus ending number K, resistance and reactance of line and the line charging admittance are included in transmission line data.

### 3.6.5    Transformer Data

The starting and ending bus number (I, K), transformer resistance and reactance and the nominal off turns ratio data must be required for each transformer connected between I and K buses.

### 3.6.6    Shunt Element Data

Shunt element data requires shunt admittance and number of bus at which shunt element is connected.

## 3.7    NEWTON RAPHSON METHOD

Newton Raphson (NR) method for load flow solution is used in this thesis because it has better convergence characteristics as compare to Golden Search (GR) method. Load flow analysis is a best way for planning of system. Golden search method is suitable for small bus system but Newton Raphson method is more reliable as compare to GS because it is useful for both small and large bus systems. NR Method has disadvantage of complex programming and it need large memory of computer. But it is still preferable because of high accuracy and lowest number of iterations. This method is useful because it can be implemented for operation of tap changing transformer, bus voltage variable constraints and optimal schedule of real and reactive. Interconnected power system is suitable as compare to isolated power system because it gives better load handling and economical operation with given reliability and security constraints. Newton Raphson method has significance to solve problem of load flow in case of interconnected power system. Optimum load flow problem's solution efficiency can be obtained using NR method, As a result line losses have been decreases by taking suitable actions like transformer tap changing position, injection of reactive power and real load sharing using matlab software [12]. Flow chart of load flow solution is shown in figure 3.3.

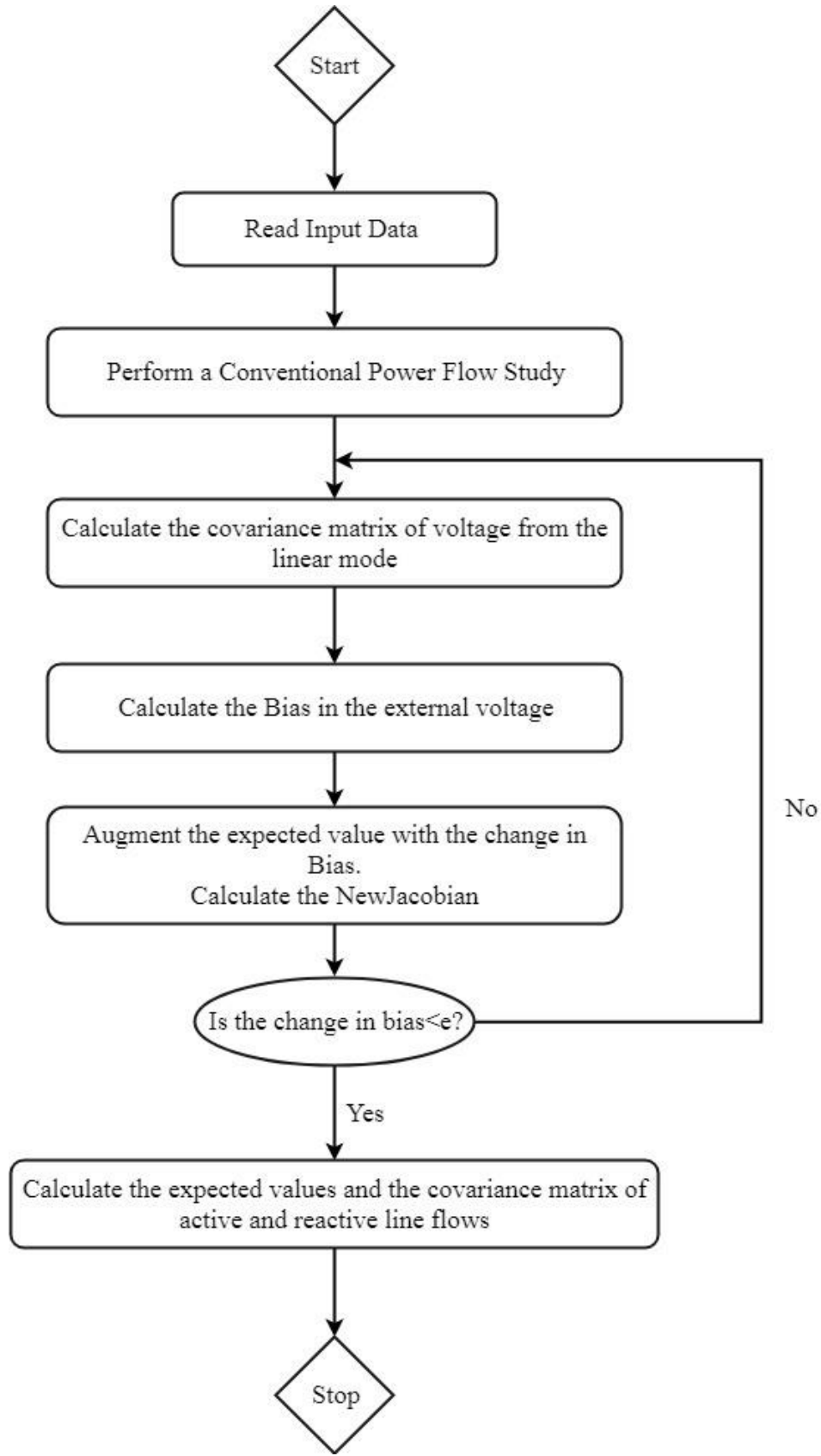

Fig. 0.3: Flow Chart of Load Flow Solution for Loss Minimization

## 3.8 GENETIC ALGORITHM

Genetic Algorithm is a technique of optimization based on Darwin's survival of hypothesis fittest. Individual animals of population are like candidate's problem solution in genetic algorithm. The initial populations are irregular or random choice of individuals. After individual's interaction, offspring produces and future generation will be formed. Weaker individuals will reproduce less often than stronger individuals. Maximization of fitness function is the objective of genetic optimization technique. For each population members, fitness is need to be calculated and best individuals are nominated for survival of next generation. Fitness value of individual is directly proportional to its probability according to roulette wheel selection.

Next generation will be formed from previous generation in the selection operation through randomly copying of chosen or selected survivors. There is possibility of copying some fit functions more than one time into the next generation causes cloning, and some functions may be not copied which are unfit causes death. There is possibility that best solution may be not selected for survival and passed over due to nature of probabilistic [60]. So, individuals having larger value of fitness will be survived and nominated to produce future better generation while lower fitness value individuals will be removed. Hence Genetic Algorithm reproduce the fittest survival between artificial chromosomes population and when there is no possibility of change in maximum value of fitness or specified number of generation is met, it stops normally [58].

The proposed GA is flexible to model and with given data of bus, system line and load forecast demand of any electrical power system, this algorithm can be implemented. Main steps of optimum power flow GA problems are following:

### 3.8.1 Generate Initial Population

The starting point of the genetic algorithm is the randomly generation of initial population. This population is composed of individuals along with different class of chromosomes.

### 3.8.2 Fitness Function

The main part of optimization problem is fitness function to determine how individuals are used to perform in the problem domain. According to problem it can be defined. In this work, there is need to minimize power loss. After evaluation of randomly generated population, objective values can be find out. Also, voltage profile can be increased through selection fitness function.

### 3.8.3    Decoding Process of Chromosomes

In power flow study, the minimization of objective function can be done through adjustment of control variable values and constraint satisfaction. The control variables may be active power generation, reactive power generation, transformer tap changing ratio, generator bus voltage the structure of chromosomes is basically an unknown vector ($Y$). The chromosome which is 5-bit binary number is used to represent the control variables value. So large number of control variables and units in power system means longer chromosomes. For generator active output power and voltage, encoding to decimal number is provided through formulation of 5-bit binary numbers. For decoding process, the formula will be following:

$$Y_i = Y_i(min) + \frac{\{Y_i(max) - Y_i(min)\}D}{(2^{BIT} - 1)} \qquad (4.22)$$

Where $Y_i$ *(min)* and $Y_i$ *(max)* are the lower and upper bound of control variables. Decoding of binary number in a gene is done in decimal number $D$. In formula, bit represents number of bits for encoding.

### 3.8.4    Parent Selection

To produce a best generation, there is need to choose best fitness values from the population as the parents. Roulette wheel selection is normally used due to its fastest convergence feature.

### 3.8.5    Crossover Operation

New offspring having inherited characteristics of its parent will be produced due to combination of two chromosomes through crossover operation. In this operation division of selected chromosomes is done into two parts and then recombination with another selected candidate is done. Other selected candidate has also been split at the same point of crossover. Single crossover point having larger crossover rate of 0.9 is selected for better performance through experiment procedure.

### 3.8.6    Mutation Operation

Mutation play important role in genetic algorithm in order to change the gene value on the string of chromosomes at random position, new discovery of genetic material, restoration of lost genetic material and for production of new genetic structure. During the run, a non-uniform mutation rate is required. Higher mutation rate is required initially to obtain large diversity. At the iteration end, small mutation rate is suitable in order to prevent destruction of good individuals which has been attained already. Mutation rate denoted as *MutR (Generation)* at corresponding number of generation can be determined using equation:

$$MutR\ (Generation) =\ MutR\ (Initial) \times e^{-\beta \times no.\ of\ Generation} \tag{4.23}$$

Where, *MutR (Initial)* is 0.9 which is initial mutation rate

$\beta$ =0.05 (a constant value)

After the evaluation of new generated chromosome's value of objective function, best offspring are added in the population which replace the weak individuals on the basis of their value of objective function. Then evaluation of fitness function is done and repetition of process occur until maximum generation is obtained.

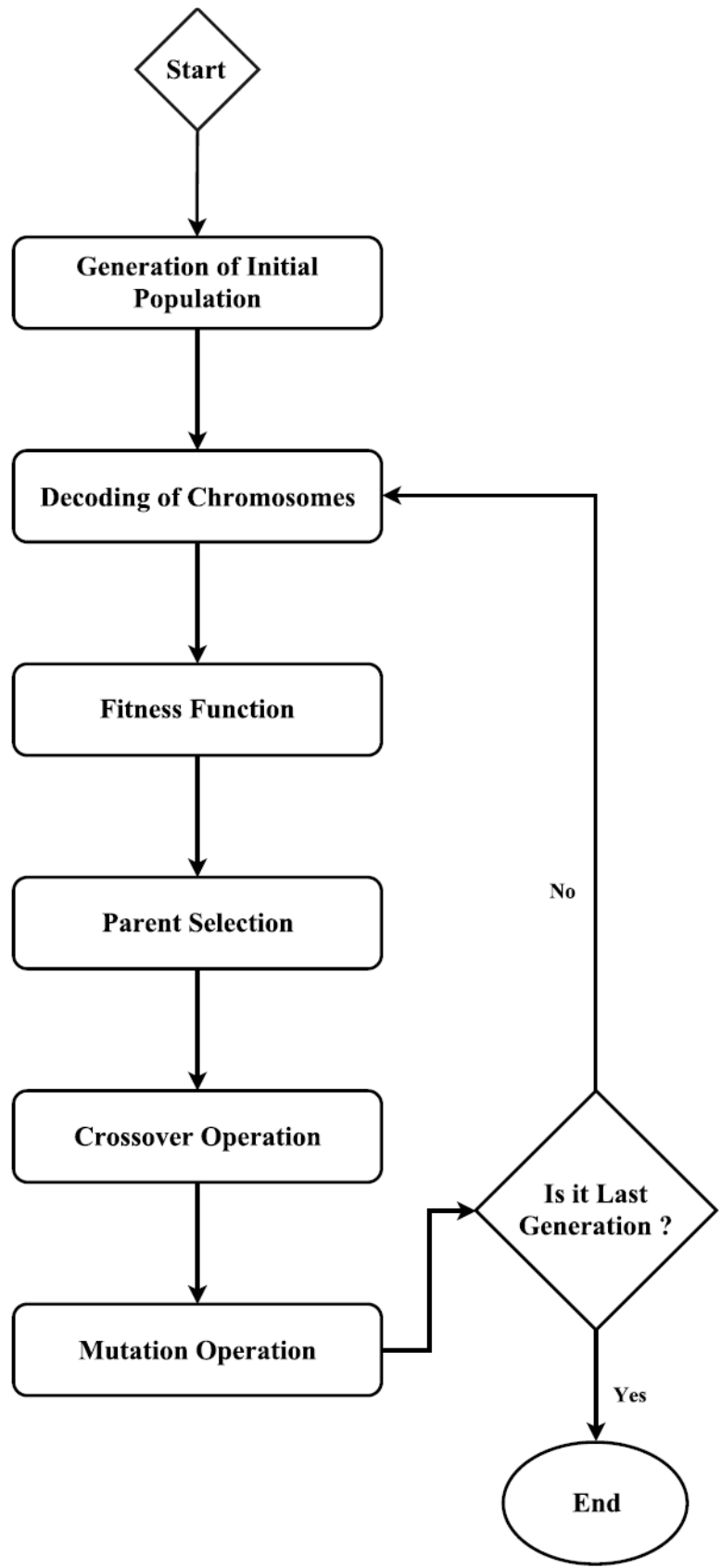

Fig. 0.4: Flow Chart of Genetic Algorithm to Minimize Line Losses

# SIMULATIONS & RESULTS

In this section, results obtained from proposed work to minimize line losses and to improve voltage profile of IEEE-30 Bus System are discussed. IEEE-30 Bus system is used as test case for optimum power flow analysis. MATLAB software is used for simulation results. The objective of work is to perform load flow analysis to improve reliability and efficiency of system by decreasing system line loses and with improvement of voltage profile.

## 4.1  STANDARD IEEE-30 BUS SYSTEM

IEEE-30 Bus system consist of 30 Buses, 6 Generators, 41 branches, 4 transformers and 21 fixed loads. Power flow analysis has been performed in IEEE-30 bus system. The bus data and transmission line data of IEEE-30 Bus system are acquired from [61]. In IEEE-30 Bus data, bus 1 indicates slack bus having voltage 1.06 p.u. with angle of 0 degree. Figure 4.1 shows IEEE-30 Bus test system.

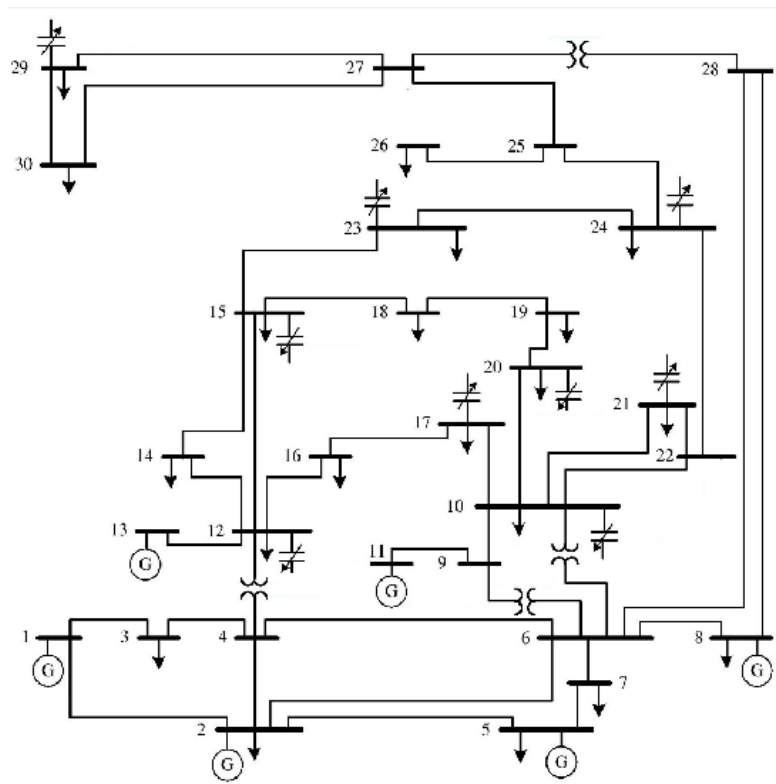

Fig. 0.1: IEEE-30 Bus System of Radial Distribution System

Regulated data. Transformer data and shunt capacitor data shown in table 4.1-4.3 are obtained from [62].

Table 0.1: Regulated Data of IEEE30 Bus

| Bus No. | Voltage Magnitude (p.u.) | Min MVAR | Max MVAR |
|---------|--------------------------|----------|----------|
| 2 | 1.043 | -40 | 50 |
| 5 | 1.01 | -40 | 40 |
| 8 | 1.01 | -10 | 40 |
| 11 | 1.082 | -6 | 24 |
| 13 | 1.071 | -6 | 24 |

Table 0.2: Transformer Data of IEEE-30 Bus

| Designation of Transformer | Per unit (p.u) Tap Setting |
|----------------------------|----------------------------|
| 4-12 | 0.932 |
| 6-9 | 0.978 |
| 6-10 | 0.969 |
| 28-27 | 0.968 |

Table 0.3: Shunt Capacitor Data

| Bus No | MVAr |
|--------|------|
| 10 | 19 |
| 24 | 4.3 |

Line losses are obtained by following strategies.

➢ **Power Flow using NR Method**

• Effect of Real load sharing on total line losses using NR

• Effect of Reactive Power Injection on total line losses using NR

• Effect of transformer tap changing on total line losses using NR

➢ **Optimum Power Flow analysis using GA**

• Comparison of Voltage profile with GA & NR

• Effect of proposed grid strategies on total Real Line Losses using GA

• Comparison of Real Line Loss of proposed grid strategies using GA & NR

## 4.2 POWER FLOW USING NR METHOD

Newton Raphson method is used in this work for load flow analysis to obtain voltage and angle profile across whole 30 buses and both real and reactive line losses are obtained. Voltage and angles across 30 buses after simulation run for base case is shown graphically in figure 4.2. Voltage across $30^{th}$ bus is 0.992p.u. which is low as compare to others. Voltage at $11^{th}$ bus is high having magnitude of 1.082p.u. Improvement of voltage profile across whole buses can reduce line losses.

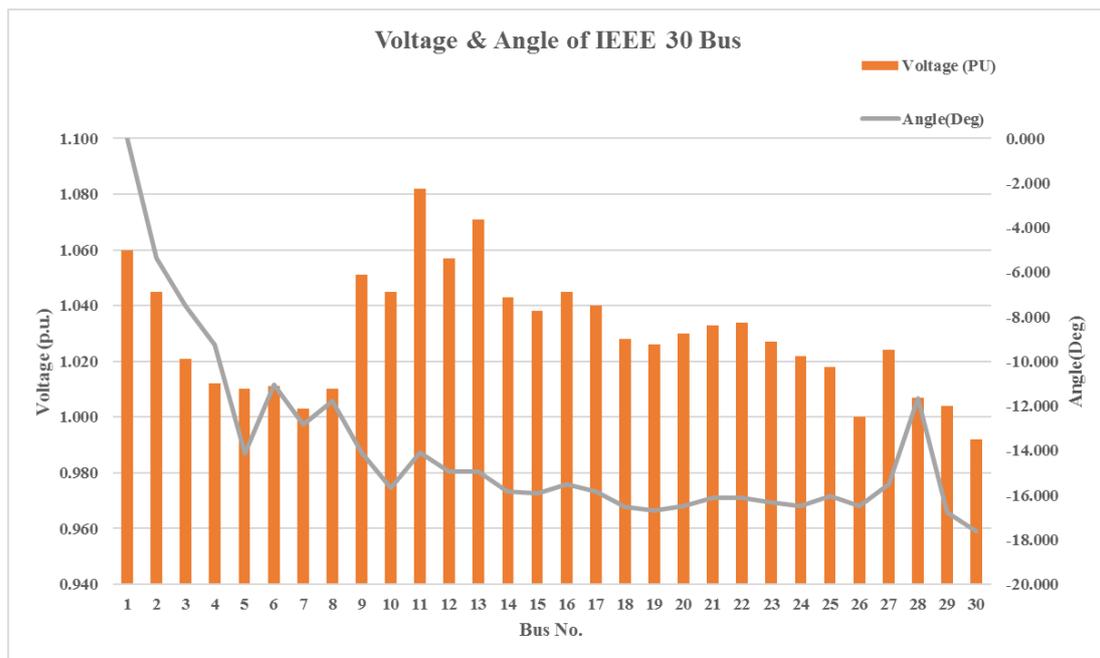

Fig. 0.2: Voltage Magnitude and Angle of IEEE 30 Bus System

Figure 4.3 shows results of line losses of IEEE-30 Bus system. Line losses includes real P(MW) and reactive Q (MVAr) line losses. Across total 41 branches both real line losses and reactive line losses are shown in below figure. These line losses are for base case without implementation of any grid strategies Total obtained line losses for base case using Newton Raphson are 17.557MW and 67.69MVAr respectively.

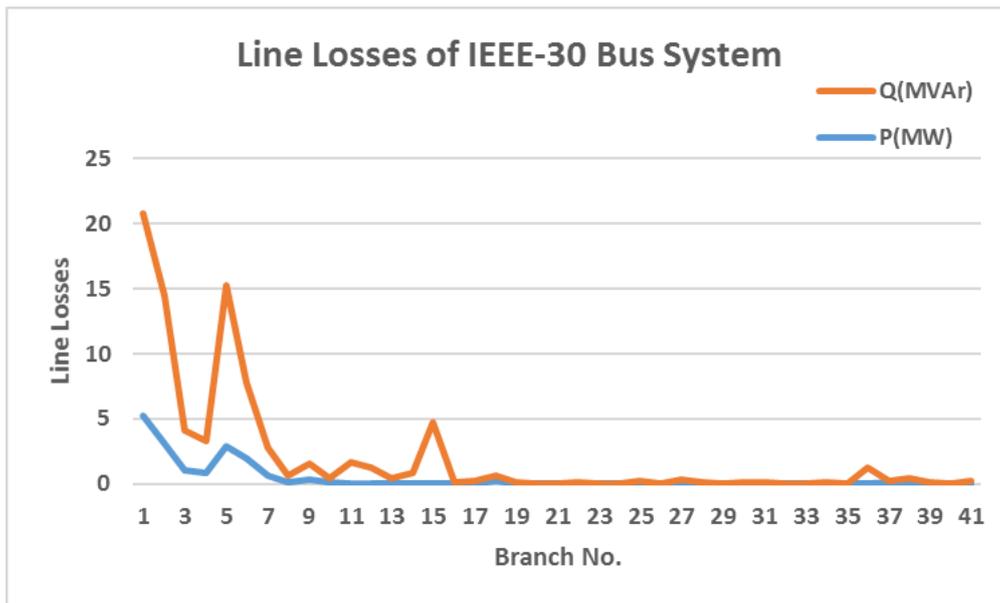

Fig. 0.3: Line Losses of IEEE-30 Bus System

Figure 4.4 shows graphical representation of total real & reactive power of generation bus, load bus and line losses for base case.

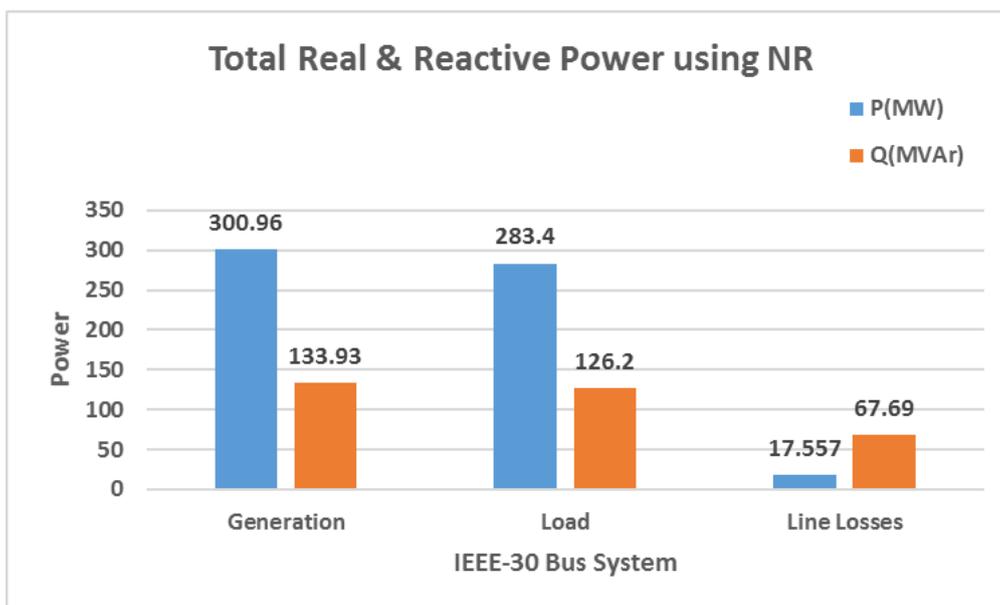

Fig. 0.4: Total Real & Reactive Power of IEEE-30 Bus System

### 4.2.1 Effect of Real Load Sharing on Line Losses using NR

Real power load sharing strategy is used to improve voltage profile and as a result line loss also reduced. In this work, 15 % real load is shared from highly loaded bus to low loaded bus. Line losses are reduced from 17.557 to 16.977 from bus high loaded bus 5 to low loaded bus 4 due to this method. Similarly, this strategy can reduce line losses from any high loaded to low loaded bus. Figure 4.5 shows real power line losses with & without load sharing and line losses are reduced by real load sharing strategy.

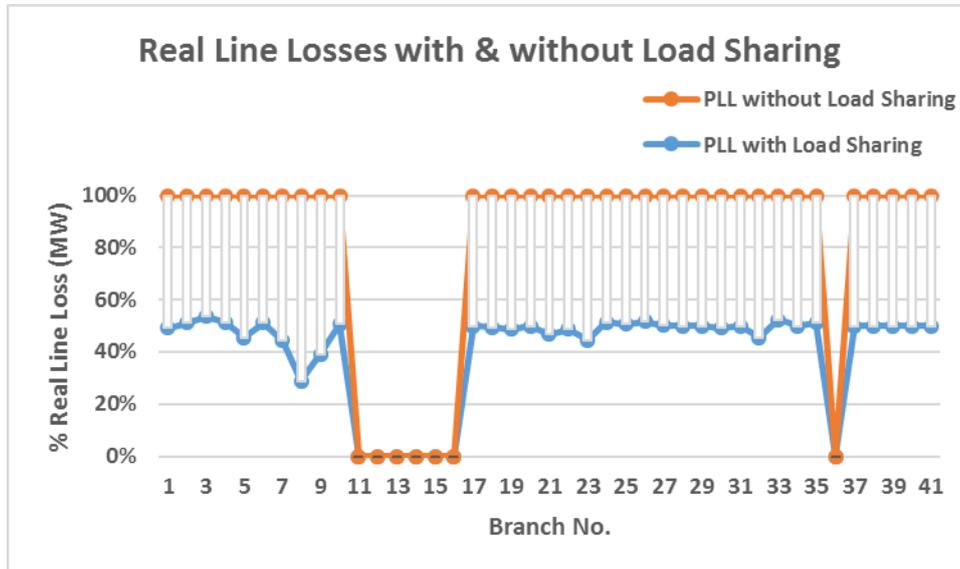

Fig. 0.5: Real Line losses with & without Load Sharing

Figure 4.6 indicates reactive power line losses with & without load sharing for comparison.

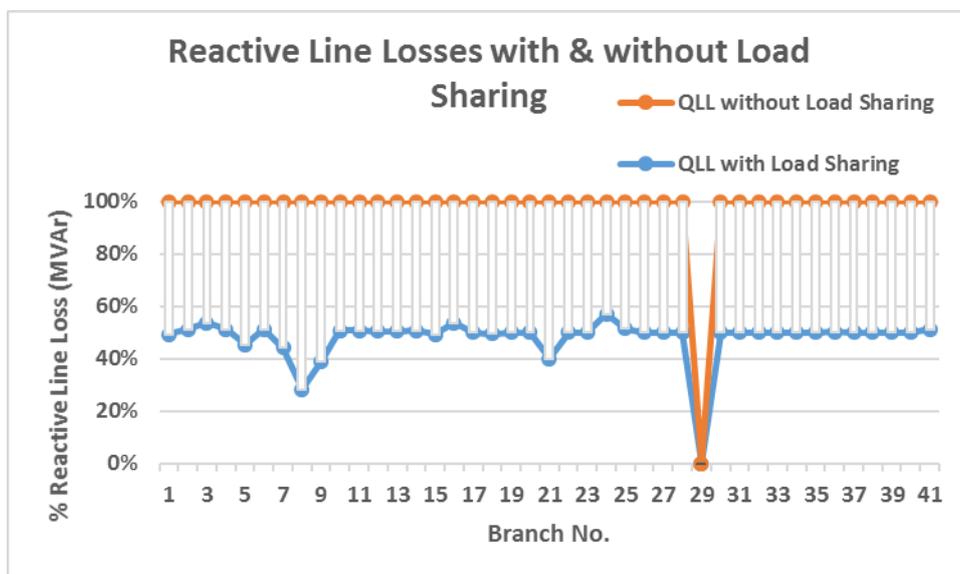

Fig. 0.6: Reactive Line losses with & without Load Sharing

Comparison of line losses for base case and with load sharing startegy is shown in figure 4.7. The graphical figure clearly indicated that due to load shsring strategy real and reactive both losses reduces significantly as compareto base case.

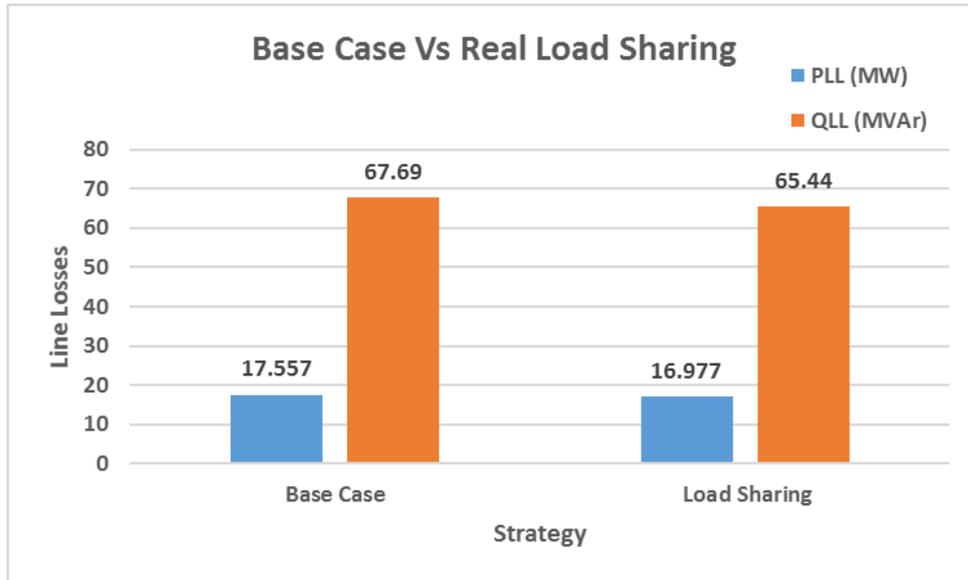

Fig. 0.7: Comparison of Base case with Load Sharing Strategy

### 4.2.2 Effect of Reactive Power Injection on Line Losses using NR

Load flow solution using NR method shows levels of voltage across load buses. Voltage across bus no.30 is low which 0.992 in comparison with to other bus voltages. Reason behind this problem is that demand of reactive power is not fulfilling from specific voltage MVAr power flow capacity of transmission line. There is need of voltage improvement across that bus. In this strategy, reactive power of 10% is injected in $30^{th}$ bus which is almost 1MVAr. Voltage is improved from 0.992to 0.999 and total line losses also reduced from 17.557MW/67.69MVAr to 17.533MW/67.59MVAr using this grid strategy. Figure 4.8 shows real power loss with and without reactive power injection strategy.

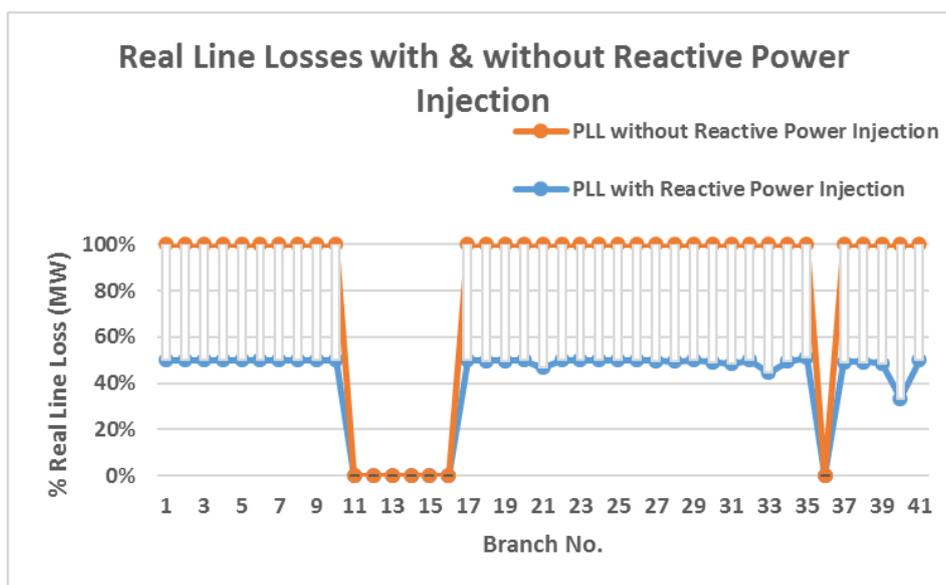

Fig. 0.8: Real Line losses with & without Reactive Power Injection

Figure 4.9 shows reactive power loss with and without reactive power injection strategy across 41 branches of IEEE-30 Bus system. Results indicates reduction of both real & reactive power losses through this grid strategy.

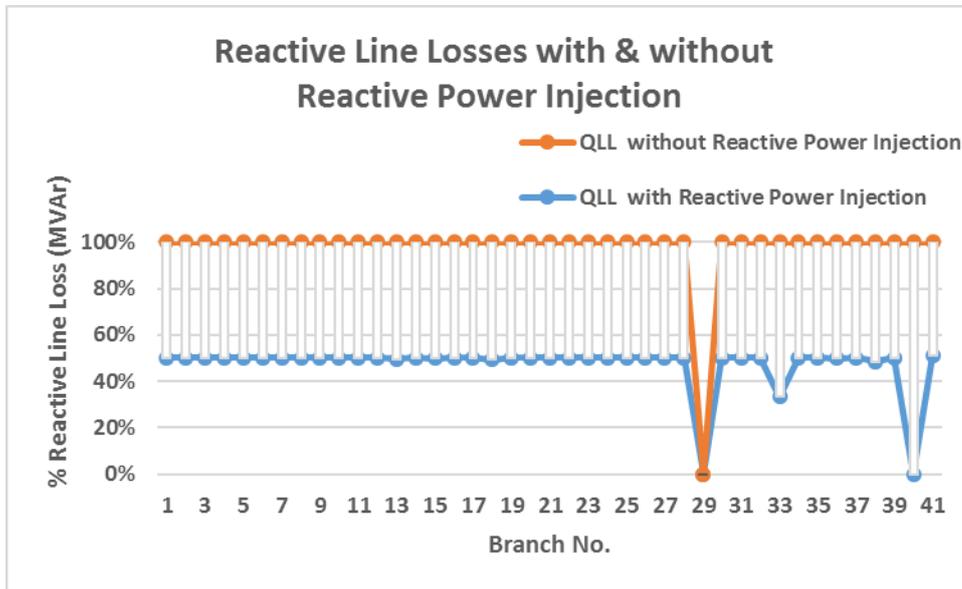

Fig. 0.9: Reactive Line Losses with & without Reactive power Injection

Fig 4.10 shows comparison between base case and both above discussed grid strategies including real load sharing and reactive power injection.

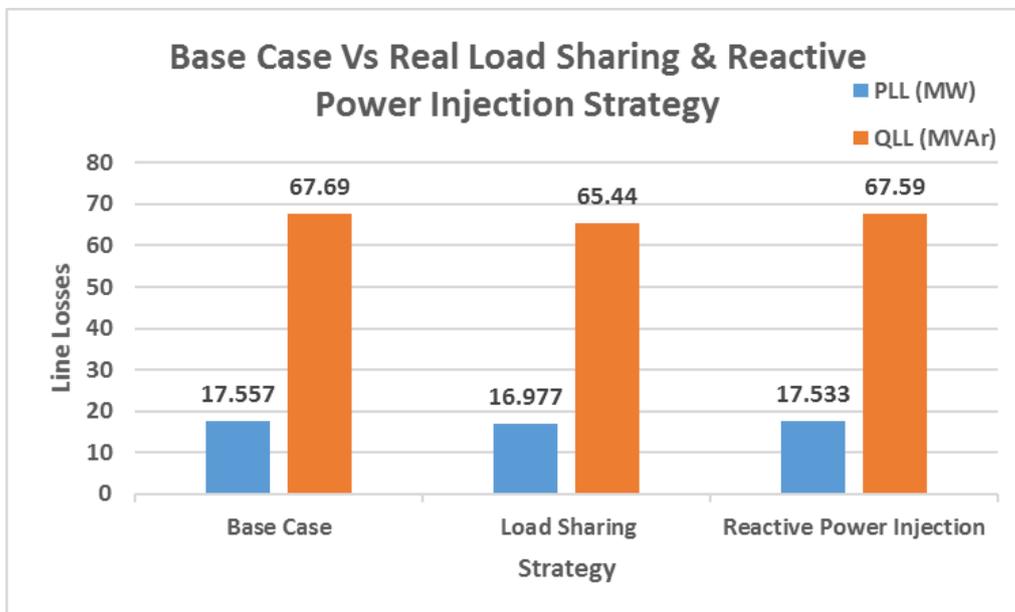

Fig. 0.10: Comparison of Base Case with Load Sharing & Reactive Power Injection Strategy

### 4.2.3 Effect of Transformer Tap Changing on Line Losses using NR

In the given data of IEEE-30 Bus, tap position of transformer at feeding bus 4 is 0. 932.With tap changing method, put 1 instead of 0.932 and after load flow analysis, it is clearly shows that total line losses reduce from 17.557MW to 17.471MW. Figure 4.11 shows Real line losses obtained with & without tap changing grid strategy.

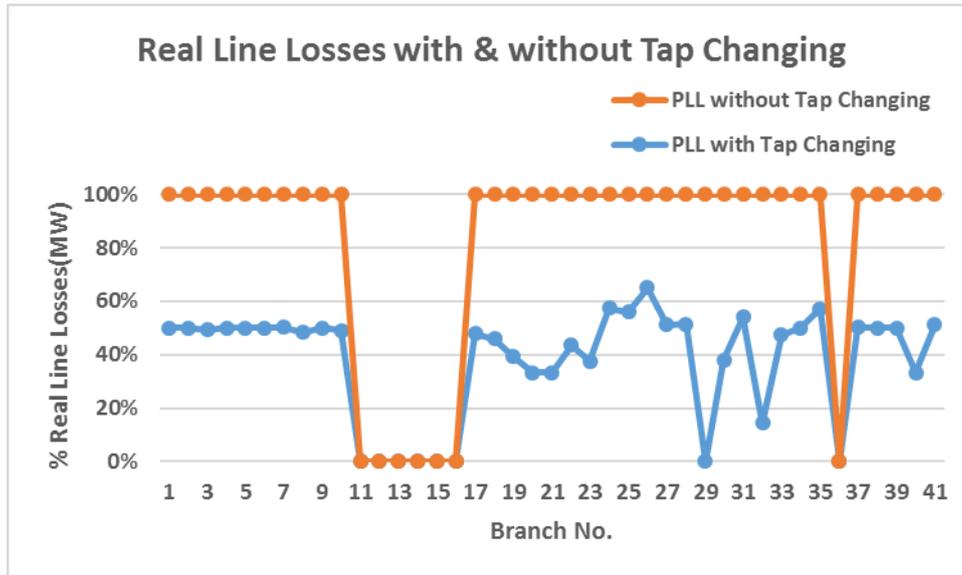

Fig. 0.11: Real Line Losses with & without Transformer Tap Changing

Reactive line losses obtained with & without tap changing grid strategy are shown in Figure 4.12 across 41 branches of IEEE-30 Bus test System.

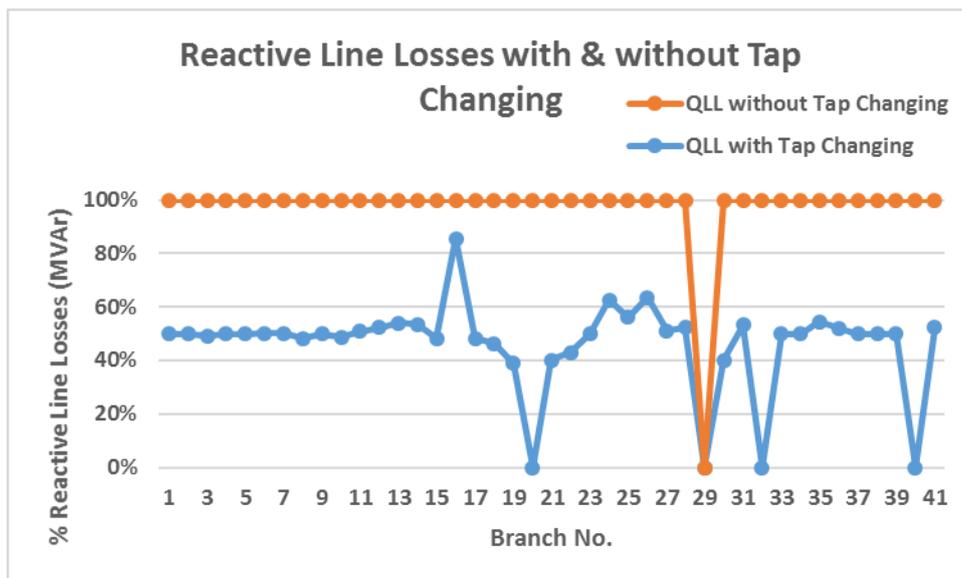

Fig. 0.12: Reactive Line Losses with & without Transformer Tap Changing

Comparison of base case with all three proposed grid strategies are shown graphically in figure 4.13. From this figure it is concluded that real line losses in base case are more as compared to all three proposed grid strategies.

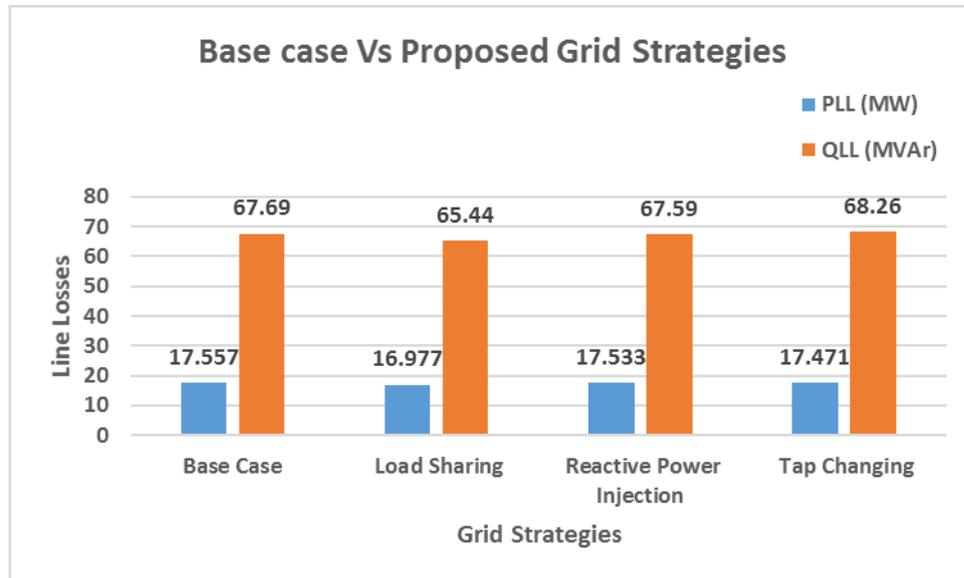

Fig. 0.13: Comparison of Base case with proposed Grid Strategies

## 4.3   OPTIMUM POWER FLOW ANALYSIS USING GA

Genetic Algorithm is second method in this work to determine line losses and voltage profile across whole 30 buses of IEEE-30 Bus system through optimum power flow analysis. Figure 4.14 show comparison between real line losses using GA and NR.

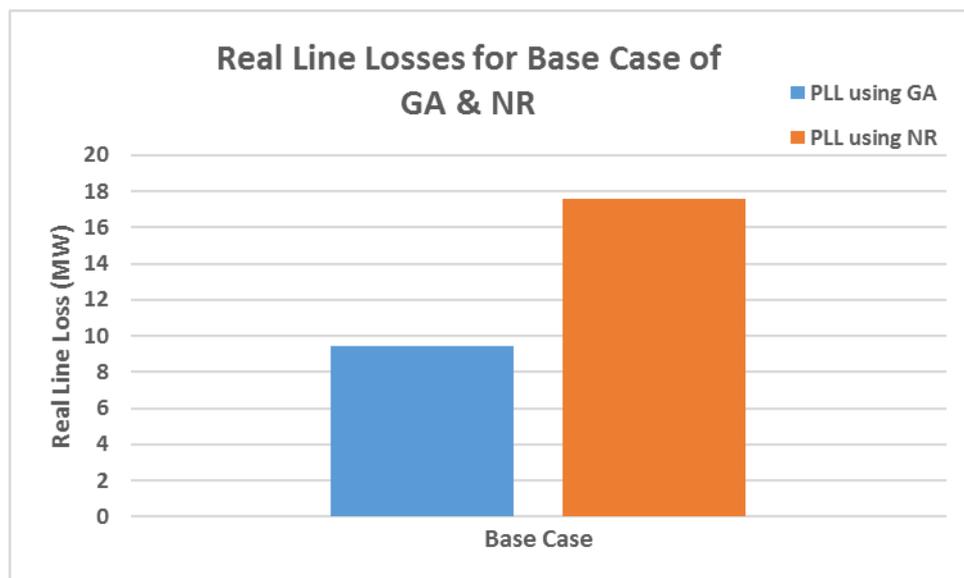

Fig. 0.14: Comparison of Real Line Losses using GA & NR

GA is more reliable and efficient method than NR with respect to line loss minimization and voltage profile improvement. In this case, 4 times simulations are run and then average is taken. Total Real Line Losses obtained using GA are 9.48145MW which are less as compared to 17.55MW real losses obtained by NR.

### 4.3.1   Comparison of Voltage Profile with GA & NR

Comparison of voltage profile using GA and NR is shown in figure 4.15. It is clearly shown from comparison that voltage is improved as compare to NR significantly across whole 30 buses of IEEE-30 bus system. Voltage across bus number 11 is more in more as compared to others using both NR & GA. Bus number 30 is low voltage bus. So voltage across whole buses of IEEE-30 Bus System must be within acceptable for reduction of real line losses.

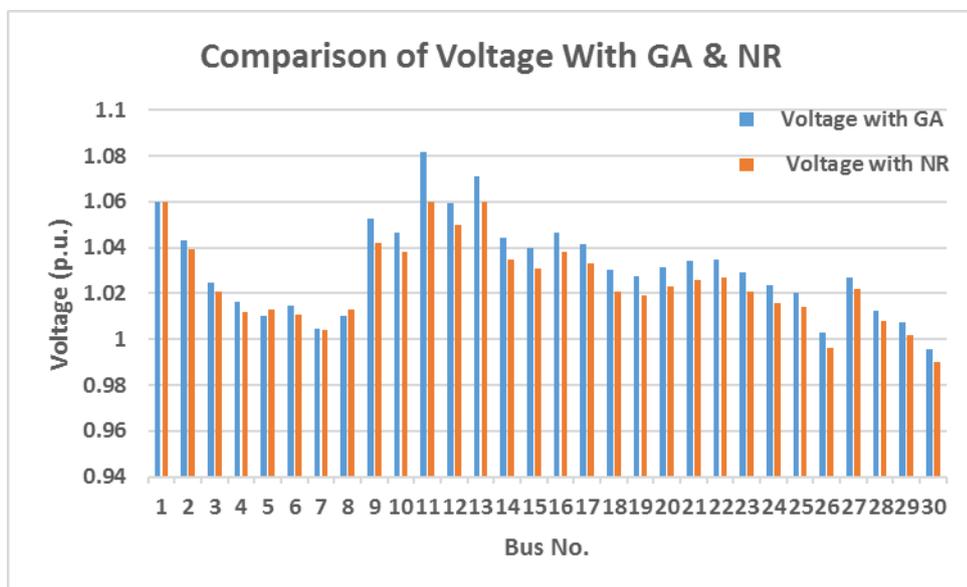

Fig. 0.15: Comparison of Voltage with GA & NR

### 4.3.2   Effect of Proposed Grid Strategies on Real Line Losses using GA

Power flow study is performed on IEEE-30 Bus system using proposed three grid strategies one by one to determine line losses. Obtained results after optimum power flow analysis are shown in table 4.4.

Table 0.4:  Real Line Loss obtained using GA

| S.No. | Base Case (MW) | Real Load Sharing (MW) | Reactive Power Injection(MW) | Transformer Tap Changing (MW) |
|---|---|---|---|---|
| 1 | 9.3213 | 9.3112 | 9.3241 | 9.698 |
| 2 | 9.6449 | 9.1936 | 9.3984 | 9.3341 |
| 3 | 9.474 | 9.3512 | 9.2683 | 8.7968 |
| 4 | 9.4857 | 9.2789 | 9.3945 | 9.5037 |
| Average | 9.481475 | 9.283725 | 9.346325 | 9.33315 |

Figure 4.16 shows line losses of base case with all three proposed grid strategies including real load sharing, reactive power injection and transformer tap changing using genetic algorithm. From average results obtained from this algorithm for each category, it is concluded that real load sharing strategy has less line losses as compared to other strategies.

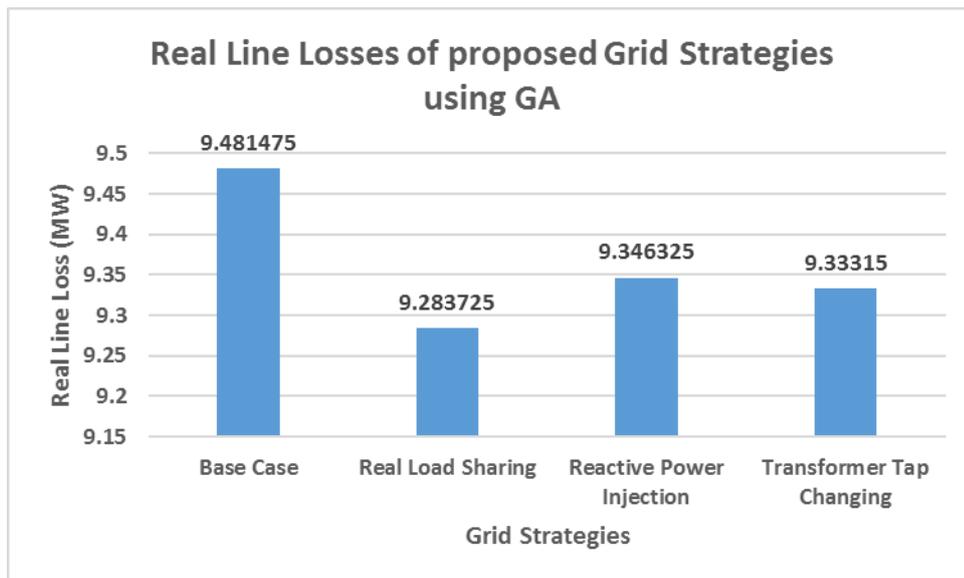

Fig. 0.16: Real Line Losses of Proposed Strategies using GA

### 4.3.3  Comparison of Real Line Loss of Proposed Grid Strategies using GA & NR

Figure 4.17 shows comparative analysis between Newton Raphson & Genetic Algorithm with respect to line losses. All three proposed grid strategies are compared with each other, with base case and with two proposed power flow methods. So obtained results shown that real line losses are minimum in case of real load sharing strategy as compare to other two strategies using both NR & GA. Transformer tap changing strategy has less line losses as compare to reactive power injection but has more losses than real load sharing. Between NR & GA, obtained results shows that genetic algorithm is reliable and efficient with respect to line losses and voltage profile improvement as compare to Newton Raphson.

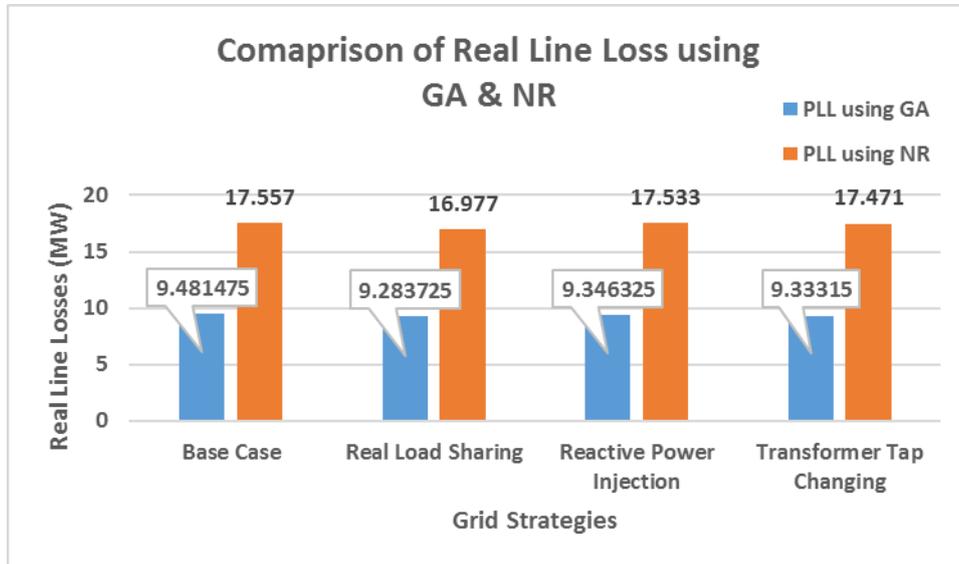

Fig. 0.17: Comparison of GA & NR with respect to Grid Strategies & Base Case

# CONCLUSION & FUTURE RECOMMENDATIONS

## 5.1 CONCLUSION

In this thesis, strategies to minimize line losses of distribution network has been demonstrated and discussed. Theses grid strategies including real load sharing, reactive power injection and transformer tap changing have been implemented on IEEE-30 Bus radial distribution system. Power flow-based study is performed using Newton Raphson & Genetic Algorithm to implement proposed grid strategies. After literature study and results obtained through simulations, it is concluded that Genetic Algorithm is efficient and reliable to minimize power line losses. As line losses has been reduced up to 9.481475MW using optimum power flow of genetic algorithm. But in base case of power flow study based on NR, real line losses are 17.557MW which are significantly very large as compare to losses obtained using GA. From grid strategies, real load sharing strategy is more suitable to minimize real line losses than other two grid strategies. Real load sharing strategy is feasible for both NR and GA as line losses are reduced from 17.557 to 16.977MW in case of NR and for GA line losses are reduced from 9.481475 to 9.23725MW. Effectiveness of the proposed scheme has been shown by obtained results.

## 5.2 FUTURE RECOMMENDATIONS

The work presented in this thesis can be enhanced in future in following directions:

- The proposed grid strategies can be implemented using PSO metaheuristic algorithm to minimize line losses
- Introduction of Distributed Generations using proposed strategies
- Towards Interconnected Distribution Network System